\documentclass[twocolumn,prd,showpacs]{revtex4}
\usepackage{graphicx}
\begin{document}

\newcommand{\re}{\mathop{\mathrm{Re}}}

\newcommand{\be}{\begin{equation}}
\newcommand{\ee}{\end{equation}}
\newcommand{\bea}{\begin{eqnarray}}
\newcommand{\eea}{\end{eqnarray}}


\title{Inhomogenized sudden future singularities}

\author{Mariusz P. D\c{a}browski}
\email{mpdabfz@sus.univ.szczecin.pl}
\affiliation{\it Institute of Physics, University of Szczecin, Wielkopolska 15,
          70-451 Szczecin, Poland}

\date{\today}

\input epsf

\begin{abstract}
We find that sudden future singularities of pressure may also appear in spatially inhomogeneous
Stephani models of the universe. They are temporal pressure singularities and may appear
independently of the spatial finite density singularities already known to exist in these
models. It is shown that the main advantage of the homogeneous
sudden future singularities which is the fulfillment of the strong
and weak energy conditions may not be the case in the inhomogeneous
case.

\end{abstract}

\pacs{98.80.Hw}
\maketitle

\section{Introduction}

It has been shown by Barrow \cite{barrow04} that for Friedmann cosmological models which
{\it do not admit} an equation of state which links the energy density $\varrho$
and the pressure $p$, a sudden future (SF) singularity of pressure may appear,
even for the matter fulfilling
the strong energy condition $\varrho > 0, \varrho + 3p > 0$, though violating the
dominant energy condition $\varrho >0,  -\varrho < p < \varrho$ \cite{lake}.
This is in contrast to the most of the current observational discussion in
cosmology, mainly devoted to determining of the barotropic index $w$
in a barotropic equation of state $p=w \varrho$, which tightly constraints
the energy density and the pressure \cite{supernovae}. On the
other hand, the observational data interpreted by such an equation
of state cannot exclude a possibility of barotropic phantom cosmological models
\cite{phantom}. These models violate null energy condition $\varrho + p > 0$,
and consequently all the remaining energy conditions \cite{he}.
Besides, phantom models allow for a Big-Rip (BR) curvature singularity,
which appears as a result of having the infinite values of the scale factor $a(t)$ at finite
future. This is in opposition to a curvature Big-Bang (BB) singularity which takes
place in the limit $a \to 0$.

The common feature of BB and BR singularities is that both $\varrho$ and $p$ blow up
equally. This is not the case with a SF singularity for which a blow up occurs only for the
pressure $p$, but not for the energy density $\varrho$. It is interesting that SF
singularities are similar to those appearing in spatially inhomogeneous Stephani models
of the universe \cite{stephani}, in
which they were termed finite density (FD) singularities \cite{sussmann,dabrowski93}.
However, FD singularities occur as singularities in spatial coordinates rather than in
time (as SF singularities do), which means that even at the present moment of the
evolution they may exist somewhere in the Universe \cite{dabrowski95,dabrowski98,chris2000}.
In this letter we show that sudden future (SF) singularities (as temporal singularities) can
be inhomogenized in the sense, that they may appear in spatially
inhomogeneous models of the universe, independently of the spatial finite density
(FD) singularities allowed in these models. We also show that
the inhomogeneous Stephani models
lead to energy conditions violation, which mainly refers to the fact that they admit
FD singularities.

The SF singularities appear in the simple framework of
Friedmann cosmology with the assumption that
the energy-momentum is conserved, so that one can write the energy density and
pressure as follows (following \cite{barrow04} we assume $8\pi
G=c=1$, $k = 0, \pm 1$)
\bea
\label{rho}
\varrho &=& 3 \left(\frac{\dot{a}^2}{a^2} + \frac{k}{a^2}
\right)~,\\
\label{p}
p &=& - \left(2 \frac{\ddot{a}}{a} + \frac{\dot{a}^2}{a^2} + \frac{k}{a^2}
\right)~.
\eea
From (\ref{p}) one is able to notice that the singularity of
pressure $p \to \infty$ occurs, when the acceleration $\ddot{a} \to -
\infty$. This can be achieved for the scale factor
\be
\label{sf2}
a(t) = A + \left(a_s - A \right) \left(\frac{t}{t_s}\right)^q -
A \left( 1 - \frac{t}{t_s} \right)^n~,
\ee
where $a_s \equiv a(t_s)$ with $t_s$ being the SF singularity time and $A, q, n =$ const.
It is obvious from (\ref{sf2}) that $a(0)=0$ and so at zero of time
a BB singularity develops.
For the sake of further considerations it is useful to write down the
derivatives of the scale factor (\ref{sf2}), i.e.,
\bea
\label{dota}
\dot{a} &=& q t_s \left(a_s - A\right)
\left(\frac{t}{t_s}\right)^{q-1} + A \frac{n}{t_s}\left( 1 - \frac{t}{t_s}
\right)^{n-1}~,\\
\label{ddota}
\ddot{a} &=& q\left(q-1\right)t_s^2\left(a_s -A \right) \left(\frac{t}{t_s}\right)^{q-2}
\nonumber \\
&-& A \frac{n(n-1)}{t_s^2}\left( 1 - \frac{t}{t_s} \right)^{n-2}~.
\eea
The main point is that the evolution of the Universe, as described
by the scale factor (\ref{sf2}), begins with the standard BB
singularity at $t=0$, and finishes at SF singularity at $t=t_s$,
provided we choose
\be
\label{nq}
1 < n < 2, \hspace{0.3cm} 0<q \leq 1~.
\ee
For these values of $n$ and $q$, the scale factor (\ref{sf2})
vanishes, and its derivatives (\ref{dota})-(\ref{ddota}) diverge
at $t=0$, leading to a divergence of $\varrho$ and $p$ in
(\ref{rho})-(\ref{p}) (BB singularity). On the other hand, the scale factor (\ref{sf2})
and its first derivative (\ref{dota}) remain constant, while its
second derivative (\ref{ddota}) diverge, leading to a divergence of
pressure in (\ref{p}) {\it only}, with finite energy density
(\ref{rho}). This behaviour means, for example, that positive curvature $(k= + 1)$
Friedmann models may not recollapse to a second BB singularity --
instead they terminate in a SF singularity \cite{BGT}.

\section{Inhomogeneized sudden future singularities}

Now, let us consider inhomogeneous Stephani models. They appear as the only
conformally flat perfect-fluid models which can be embedded in a 5-dimensional flat
space \cite{stephani,krasinski}. Their metric in the spherically symmetric case reads as
(notice that we have introduced a Friedmann-like time coordinate which
eliminated one of the functions of time in the original Stephani
metric \cite{dabrowski93})
\bea
\label{STMET}
ds^2~&=&~-~\frac{a^2}{\dot{a}^2} \frac{a^2}{V^2}
\left[\left( \frac{V}{a} \right)^{\cdot} \right]^2 dt^2~ \nonumber \\
&+&~ \frac{a^2}{V^2} \left[dr^2~+~r^2 \left(d\theta^2~+~
\sin^2{\theta}d\varphi^2 \right) \right]~,
\eea
where
\be
\label{VSS}
  V(t,r)  =  1 + \frac{1}{4}k(t)r^2~,
\ee
and $(\ldots)^{\cdot}~\equiv~\partial/\partial t$. The function $a(t)$
plays the role of a generalized scale factor, $k(t)$ has the meaning of a
time-dependent "curvature index", and $r$ is the radial coordinate.

Their analogy to SF singularity models is that they do not admit {\it any} equation of
state linking $p$ to $\varrho$ throughout the whole evolution of the universe, although at
any given moment of the evolution, such an equation of state
(varying from one spacelike hypersurface to the other) can be admitted.
An analytic equation of state can also be admitted
at any fixed subspace with constant radial coordinate $r$, but not globally
\cite{dabrowski93}. For the sake of simplicity, first the spherically symmetric
models will be considered (note that
in \cite{dabrowski93,dabrowski95} a time coordinate $t$ analogous to the cosmic time
in Friedmann models was marked by $\tau$, which had nothing to do with a common
conformal time coordinate). The energy density and pressure are given by
\cite{dabrowski93}
\begin{eqnarray}
\label{rhost}
\varrho(t) & = & 3C^2(t) \equiv 3 \left[ \frac{\dot{a}^2(t)}{a^2(t)} + \frac{k(t)}{a^2(t)}
\right]~,\\
\label{pst}
p(t,r) & = & -~3C^2(t)~+~2C(t) \dot{C}(t) \frac{ \left[ \frac{V(t,r)}{a(t)} \right]}
  { \left[ \frac{V(t,r)}{a(t)} \right]^{\cdot}}~,
\end{eqnarray}
and generalize the relations (\ref{rho}) and (\ref{p}) to
inhomogeneous models.

We now show that it is possible to extend SF singularities into
inhomogeneous models. Following \cite{dabrowski93} we assume that
the functions $k(t)$ and $a(t)$ are
related by
\be
\label{kaba}
k(t) = - \alpha a(t)~,
\ee
with $\alpha =$ const. In fact, the limit $\alpha \to 0$
gives the Friedmann models (cf. the discussion of the conditions
to derive such a limit in \cite{dabrowski93}). Inserting
(\ref{kaba}) into (\ref{pst}) we get
\bea
\label{rhokaba}
\varrho(t) & = & 3 \left[ \frac{\dot{a}^2(t)}{a^2(t)} - \frac{\alpha}{a(t)}
\right]~, \\
\label{pkaba}
p(t,r) &=& - 2 \frac{\ddot{a}}{a} - \frac{\dot{a}^2}{a^2}
+ 2 \frac{\alpha}{a} - \frac{1}{4} \alpha r^2 \left( 2 \frac{\dot{a}^2}{a} - 2 \ddot{a}
-  \alpha \right)~.
\eea
From (\ref{pkaba}) one can see, that $p \to \infty$, when acceleration $\ddot{a} \to -
\infty$ for an {\it arbitrary} value of the radial coordinate $r$. We
can then say that we {\it generalized} SF singularities (given, for example, by the scale factor
(\ref{sf2})) onto an inhomogeneous model of the universe.

However, it is very interesting to notice that in such a
generalization not only SF singularities appear, but also FD
singularities are possible for the radial coordinate $r^2 \to
\infty$. These seem to be far away from us, and so not very harmful,
since $r^2 \to \infty$ defines an antipodal center of symmetry in the spherically
symmetric Stephani models. The advantage of these FD singularities is that they are able to
drive the current acceleration of the universe \cite{dabrowski98,stelmach01,dabrowski02}.

The procedure of inhomogenizing SF singularities may be extended into the general Stephani
models for which there is no spacetime symmetry at all, and so they are completely
inhomogeneous. The most general Stephani metric in cartesian coordinates
$(x,y,z)$ \cite{stephani,krasinski} reads as
\bea
\label{STMETGEN}
ds^2~&=&~-~\frac{a^2}{\dot{a}^2} \frac{a^2}{V^2}
\left[\left( \frac{V}{a} \right)^{\cdot} \right]^2 dt^2~ \nonumber \\
&+&~ \frac{a^2}{V^2} \left[dx^2~+dy^2~+~dz^2 \right]~,
\eea
where
\bea
\label{Vtxyz}
&& V(t,x,y,z) = \\
&& 1 + \frac{1}{4} k(t) \left\{ \left[x -
x_0(t) \right]^2 + \left[y - y_0(t) \right]^2 + \left[z - z_0(t)
\right]^2 \right\}~,\nonumber
\eea
and $x_0, y_0, z_0$ are arbitrary functions of time.
Now the general expression for the pressure is (the expression for the energy
density (\ref{rhost}) remains the same)
\be
\label{pgen1}
p(t,x,y,z)  =  -~3C^2(t)~+~2C(t) \dot{C}(t) \frac{ \left[ \frac{V(t,x,y,z)}{a(t)} \right]}
  { \left[ \frac{V(t,x,y,z)}{a(t)} \right]^{\cdot}}~.
\ee

Inserting the time derivative of (\ref{rhost}) and the function $V(t,x,y,z)$ from
(\ref{Vtxyz}) into (\ref{pgen1}) gives
\bea
\label{pgen2}
&&p(t,x,y,z)  =  -~3\frac{\dot{a}^2}{a^2} - 3\frac{k}{a^2}~
\\
&&+ \frac{\dot{a}}{a} \left[ 2 \frac{\ddot{a}}{a} - 2 \frac{\dot{a}^2}{a^2}
+ \frac{1}{a^2} \left(\dot{k} \frac{a}{\dot{a}} - 2 k \right) \right]
\frac{ \left[ \frac{V(t,x,y,z)}{a(t)} \right]}
  { \left[ \frac{V(t,x,y,z)}{a(t)} \right]^{\cdot}}~\nonumber .
\eea
It is easy to notice that SF singularity $p \to \pm\infty$ appears
with (\ref{sf2}) for $\ddot{a} \to -\infty$, if $(V/a)/(V/a)^{\cdot}$
is regular and the sign of the pressure depends on the signs of both $\dot{a}/a$
and $(V/a)/(V/a)^{\cdot}$. This proves that we can {\it inhomogenize} SF
singularities for a Stephani model with no symmetry.

In fact, SF singularities appear independently of FD singularities
whenever $\ddot{a} \to -\infty$ and the blow-up of $p$
is guaranteed by the involvement of the time derivative of the
function $C(t)$ in (\ref{pst}).

That makes a complimentary generalization to the one which extends
SF singularities into the theories with actions being arbitrary
analytic functions of the Ricci scalar and into
anisotropic (but homogeneous) models \cite{barrow041,barrow042,abdalla04}.

It appears that the main motivation for studying SF singularities \cite{barrow04}
was the fact that, unlike phantom models \cite{phantom}, they obey
the strong and weak energy conditions, though they do not obey the
dominant energy condition. In this paper we raise the point that
the question of possible violation of the energy conditions for
the inhomogenized SF singularity models is a bit more complex than for
the homogeneous ones. From (\ref{pst}) and (\ref{pkaba}) for the strong, weak and
dominant energy conditions to be fulfilled we have:
\bea
\label{en1}
\varrho + 3p &=& - 6 \frac{\ddot{a}}{a} + 3 \frac{\alpha}{a}
\nonumber \\
&-&
\frac{3}{4} \alpha r^2 \left( 2 \frac{\dot{a}^2}{a} - 2 \ddot{a}
-  \alpha \right) > 0~,\\
\label{en2}
\varrho + p &=& - 2 \frac{\ddot{a}}{a} + 2 \frac{\dot{a}^2}{a^2}
-  \frac{\alpha}{a} \nonumber \\
&-& \frac{1}{4} \alpha r^2 \left( 2 \frac{\dot{a}^2}{a} - 2 \ddot{a}
-  \alpha \right) > 0~,\\
\label{en3}
\varrho - p &=& 2 \frac{\ddot{a}}{a} + 4 \frac{\dot{a}^2}{a^2}
- 5 \frac{\alpha}{a} \nonumber \\
&+& \frac{1}{4} \alpha r^2 \left( 2 \frac{\dot{a}^2}{a} - 2 \ddot{a}
-  \alpha \right) > 0~.
\eea
In fact, the dominant energy condition requires fulfilling both
(\ref{en2}) and (\ref{en3}). Notice that in view of (\ref{rhost})
the energy density in inhomogeneous Stephani models is always positive, i.e.,
\be
\label{en4}
\varrho > 0~.
\ee
This means that the strong and weak energy conditions are still
not violated if $\ddot{a} \to -\infty$, in analogy to homogeneous
models, provided
\be
\label{over}
\frac{1}{a} > \frac{\alpha}{4} r^2~,
\ee
since the first term with $\ddot{a}$ in (\ref{en1}) and
(\ref{en2}) must dominate the second (remember that $\ddot{a} \to -\infty$).
Notice that the equality $1/a = \alpha r^2/4$
may lead to a pressure singularity avoidance in (\ref{pkaba}).
Assuming that the generalized scale factor $a(t)>0$, we conclude from (\ref{over})
that the strong and weak energy conditions are always fulfilled, if $\alpha <0$. However,
an accelerated expansion for an observer at the center of symmetry at $r=0$
can only be achieved, if $\alpha > 0$
\cite{dabrowski02} (the spatial
acceleration scalar reads as $\dot{u} = - 2 \alpha r$ -- lower pressure regions
are away from the center). This means that the strong and weak
energy conditions are {\it not necessarily fulfilled} for the models
with $\alpha >0$. In particular, they cannot be fulfilled at an
antipodal center of symmetry at $r^2 \to \infty$, unless $a \to 0$
(where Big-Bang singularity appears and so BB and FD singularities
coincide - see Section \ref{FD}).

On the other hand,
the first part of the dominant energy condition may not be violated if
the contribution from the last term with $\ddot{a}$ in (\ref{en3}), which includes
$r^2$, does not overweigh the first one, i.e., when
\be
\label{over1}
\frac{1}{a} < \frac{\alpha}{4} r^2~.
\ee
This should be appended by the condition (\ref{en2}) which is
equivalent to (\ref{over}), i.e., the dominant energy condition is
fulfilled if
\be
\label{over2}
\frac{1}{a} < \frac{\alpha}{4} r^2 < \frac{1}{a}~.
\ee
This is obviously a contradiction which means that, similarly as in
the isotropic Friedmann models, SF singularities violate the dominant
energy condition.

Such a violation of the dominant energy condition also
appears in M-theory-motivated ekpyrotic models in which $p \gg \varrho$
during recollapse \cite{turok}.

Let us now discuss the problem of the possible energy conditions
violation in the general Stephani model. Using (\ref{rhost}) and
(\ref{pgen2}) we get for the strong, weak, and dominant energy
conditions
\bea
\label{strgen}
&&\varrho + 3p  =  -~6\frac{\dot{a}^2}{a^2} - 6\frac{k}{a^2}~ +
\\
&&3 \frac{\dot{a}}{a} \left[ 2 \frac{\ddot{a}}{a} - 2 \frac{\dot{a}^2}{a^2}
+ \frac{1}{a^2} \left(\dot{k} \frac{a}{\dot{a}} - 2 k \right) \right]
\frac{ \left[ \frac{V(t,x,y,z)}{a(t)} \right]}
  { \left[ \frac{V(t,x,y,z)}{a(t)} \right]^{\cdot}} >
  0~,\nonumber\\
\label{weakgen}
&&\varrho + p  = \\
&&2 \frac{\dot{a}}{a} \left[ 2 \frac{\ddot{a}}{a} - 2 \frac{\dot{a}^2}{a^2}
+ \frac{1}{a^2} \left(\dot{k} \frac{a}{\dot{a}} - 2 k \right) \right]
\frac{ \left[ \frac{V(t,x,y,z)}{a(t)} \right]}
  { \left[ \frac{V(t,x,y,z)}{a(t)} \right]^{\cdot}} >
  0~,\nonumber\\
\label{domgen}
&&\varrho - p  = 6\frac{\dot{a}^2}{a^2} + 6\frac{k}{a^2} \\
&& - \frac{\dot{a}}{a} \left[ 2 \frac{\ddot{a}}{a} - 2 \frac{\dot{a}^2}{a^2}
+ \frac{1}{a^2} \left(\dot{k} \frac{a}{\dot{a}} - 2 k \right) \right]
\frac{ \left[ \frac{V(t,x,y,z)}{a(t)} \right]}
  { \left[ \frac{V(t,x,y,z)}{a(t)} \right]^{\cdot}} >
  0~.\nonumber
\eea
Obviously, the dominant energy condition requires fulfilling both (\ref{weakgen}) and
(\ref{domgen}). Before we go any further, using
(\ref{Vtxyz}), we note that
\bea
\label{Va}
&& \frac{V(t,x,y,z)}{a(t)} = \\
&& \frac{1}{a} + \frac{1}{4} \frac{k}{a} \left[ \left(x -
x_0 \right)^2 + \left(y - y_0 \right)^2 + \left(z - z_0
\right)^2 \right]~, \nonumber \\
\label{Vadot}
&& \left[ \frac{V(t,x,y,z)}{a(t)} \right]^{\cdot} = \\
&& -\frac{\dot{a}}{a^2} + \frac{1}{4a} \left( \dot{k} -
k \frac{\dot{a}}{a} \right) \left[ \left(x -
x_0 \right)^2 + \left(y - y_0 \right)^2 + \left(z - z_0
\right)^2 \right] \nonumber \\
&& - \frac{k}{2a} \left[ \left(x -
x_0\right)\dot{x}_0 +  \left(y -
y_0\right)\dot{y}_0 +  \left(z -
z_0\right)\dot{z}_0 \right]~. \nonumber
\eea
It is important to notice that the ratio of (\ref{Va}) and (\ref{Vadot})
which appears in the conditions (\ref{strgen}), (\ref{weakgen}),
and (\ref{domgen}) allows to cancel $1/a$ from both the
numerator and the denominator. Apart from that, one is able to
take $\dot{a}/a$ out in (\ref{Vadot}) and cancel it with the same
term standing in front of the last term in these conditions. This
basically means that, bearing in mind the fact that $\ddot{a} \to
- \infty$, the strong and weak energy conditions are fulfilled
provided one of the expressions
\bea
\label{Vac}
&& V_1 \equiv 1 + \frac{1}{4} k \left[ \left(x -
x_0 \right)^2 + \left(y - y_0 \right)^2 + \left(z - z_0
\right)^2 \right]~, \\
\label{Vadotc}
&& V_2 \equiv  \frac{1}{4} \left( \dot{k}\frac{a}{\dot{a}} - k  \right) \left[ \left(x -
x_0 \right)^2 + \left(y - y_0 \right)^2 + \left(z - z_0
\right)^2 \right] \nonumber \\
&& - \frac{ka}{\dot{a}} \left[ \left(x -
x_0\right)\dot{x}_0 +  \left(y -
y_0\right)\dot{y}_0 +  \left(z -
z_0\right)\dot{z}_0 \right]~ - 1,
\eea
is negative. It is clear that for $k(t)>0$ (\ref{Vac}) is always
positive, so that (\ref{Vadotc}) must necessarily be negative and
it certainly does, at least in some regions of space. On the other
hand, for $k(t)<0$ (\ref{Vac}) can be both positive and negative
which requires (\ref{Vadotc}) to be negative and positive,
respectively. In conclusion, similarly as in the spherically
symmetric case, the strong and weak energy conditions {\it may be
violated} for inhomogenized  SF singularities in Stephani models. This
is different from what we have for isotropic SF singularities.
Finally, one can easily notice that in order to fulfill the
dominant energy condition the ratio of (\ref{Vac}) to
(\ref{Vadotc}) should be simultaneously positive and negative,
which is a contradiction. This means that, like in the isotropic
case, a general Stephani model allows SF singularities which
{\it always violate} the dominant energy condition.
In conclusion, one can say that the problem of the energy
conditions violation by SF singularities in the inhomogeneous models
is more complicated than in the isotropic ones and so the results
based only on the isotropic models cannot be trusted.

\section{Inhomogeneous finite density singularities}
\label{FD}

In the context of temporal SF singularities of pressure we will further briefly discuss the
occurrence of spatial FD singularities of pressure in the Stephani models and possible energy
conditions violation. From (\ref{pst}) we can see that FD
singularities appear whenever the radial coordinate
\be
r^2 = - 4 \frac{ \left( \frac{1}{a} \right)^{\cdot}}
  { \left( \frac{k}{a} \right)^{\cdot}}~.
\ee
Under the choice of (\ref{kaba}), we have $(k/a)^{\cdot}=0$, and so
the singularities appear at $r^2 \to \infty$. In general, it may
not be so, which was explicitly shown in \cite{dabrowski93}. For
example, by choosing
\bea
\label{adelta}
a(t) = \alpha t^2 + \beta t + \gamma~,\\
\label{kabadelta}
k(t) = 1 - \dot{a}^2 = - 4 \alpha a(t) + \Delta~,
\eea
with
\be
\Delta = 4\alpha\gamma + 1 - \beta^2~,
\ee
the FD singularities appear for \cite{dabrowski93}
\be
|r| = 2/\sqrt{-\Delta}~.
\ee
Of course the condition (\ref{kaba}) is obtained in the limit $\Delta \to 0$
from (\ref{kabadelta}) which moves FD singularities to an antipodal
center of symmetry at $r^2 \to \infty$. Having chosen $\gamma=0, \Delta = 1 - \beta^2$
in (\ref{adelta}) (Model I of Ref. \cite{dabrowski98}, called D\c{a}browski model in
Ref. \cite{chris2000}) the energy density and pressure are then given by
\bea
\varrho &=& 3 \frac{1}{t^2(\alpha t + \beta)^2}~,\\
p &=& - \frac{1 + 2\alpha t (\alpha t + \beta) r^2}{t^2(\alpha
t + \beta)^2}~.
\eea
For the strong, weak and dominant energy conditions to be fulfilled, respectively,
we get the requirements
\bea
\varrho + 3p &=& - 6 \alpha \frac{r^2}{t(\alpha t + \beta)} > 0~,\\
\varrho + p &=& 2 \frac{1 - \alpha t (\alpha t + \beta) r^2}{t^2(\alpha
t + \beta)^2} > 0~,\\
\varrho - p &=& 2 \frac{2 + \alpha t (\alpha t + \beta) r^2}{t^2(\alpha
t + \beta)^2} > 0~.
\eea
If $\alpha>0$ (acceleration \cite{dabrowski02}), then the strong energy
condition is violated if
\be
\label{strongc}
t(\alpha t + \beta) >0~,
\ee
and this may happen independently of the
radial coordinate $r$. The weak energy condition is violated for
the domain of space in which
\be
\label{weakc}
\frac{1}{r^2} > \alpha t(\alpha t + \beta)~.
\ee
With the strong energy condition violated, this gives a weak energy
condition violation only in some spatial domain since the right-hand side of
(\ref{weakc}) has a positive value, but including the center of symmetry at $r=0$.
On the other hand, for the decelerated expansion, $\alpha <0$, and
the right-hand side of (\ref{weakc}) has a negative value, so the
weak energy condition is violated everywhere in the universe (i.e., for
all values of $r$). If the strong energy condition is not violated
and $\alpha>0$ (acceleration), then again there is a weak energy
condition violation for all values of $r$. If the strong energy
condition is violated, and $\alpha<0$, then the weak energy condition
is violated only in some spatial domain which, however, includes
the center of symmetry at $r=0$. The dominant energy condition is violated for
\be
\label{domc}
\frac{1}{r^2} < - \frac{1}{2} \alpha t(\alpha t + \beta)~,
\ee
which, combined with (\ref{weakc}), gives
\be
\label{domcfull}
\alpha t(\alpha t + \beta) < \frac{1}{r^2} < - \frac{1}{2} \alpha t(\alpha t + \beta)~.
\ee
This last condition can only be fulfilled either if $\alpha <
0$ and $t(\alpha t + \beta) >0$, or if $\alpha >
0$ and $t(\alpha t + \beta) <0$. Then, at least for these particular class
of Stephani models, FD singularities may lead to a violation of the energy
conditions in a similar way as BR singularities in phantom
cosmology do.

An interesting problem is a possible avoidance of SF and FD
singularities in the universe. SF singularities can easily be
avoided by imposing an analytic form of the equation of state
$p=p(\varrho)$. Even without this assumption, some other ways of their avoidance
by introducing quadratic in Ricci curvature scalar terms
\cite{abdalla04}, or by quantum effects \cite{nojiri04}, are
possible. On the other hand, a necessary condition to avoid FD
singularities in Stephani models comes from (\ref{pst}) and reads
as
\be
\frac{ \left( \frac{1}{a} \right)^{\cdot}}
  { \left( \frac{k}{a} \right)^{\cdot}} > 0~.
\ee
In our special model (\ref{adelta})-(\ref{kabadelta}) they can be
simply avoided, if
\be
\Delta > 0~.
\ee

\section{Conclusions}

In conclusion, we have shown that one is able to spatially inhomogenize
sudden future (SF) singularities in the sense that
these singularities {\it do appear} in inhomogeneous models of
the universe. However, despite homogeneous SF singularities, they
{\it may} violate the strong and weak energy conditions in some regions
of space, although they {\it share} the dominant energy condition violation with
homogeneous models. It shows that making any important conclusions about
physics, on the basis of the isotropic models only, may be misleading and
should not be trusted. A possible violation of all the energy
conditions by inhomogenized SF singularities is similar to what
happens to Big-Rip singularities in phantom cosmologies.

Besides, we
have noticed that, apart from sudden future singularities, the inhomogenized models
also admit finite density (FD) singularities which are spatial rather
than temporal. In relation to this we have discussed an example of an
inhomogeneous model with spatial finite density singularities of
pressure and studied the domains of its energy conditions
violation.

\section{Acknowledgements}

I acknowledge the support from the Polish Research Committee (KBN)
grant No 2P03B 090 23.

\end{document}